The Structural, Energetic and Electronic Properties of Doped Carbon Nanotubes by Encapsulation of MCp$_2$ (M=Fe, Co, Ni): a Theoretical Investigation

*Fenglei Cao [1,2], Wei Ren [3*], Xianfang Xu [1], Ye-Xiang Tong [1], and Cunyuan Zhao [1*]*

[1]MOE Laboratory of Bioinorganic and Synthetic Chemistry, School of Chemistry and Chemical Engineering, Sun Yat-sen University, Guangzhou 510275, China

[2]School of Chemistry and Chemical Engineering, Shanghai Jiao Tong University, Shanghai 200240, China

[3]Physics Department, University of Arkansas, Fayetteville, Arkansas 72701, USA

*E-mail: weiren@uark.edu     *E-mail: ceszhcy@mail.sysu.edu.cn

**ABSTRACT**

Metallocenes can be encapsulated inside the carbon nanotubes. The structural, energetic and electronic properties of organometallic $MCp_2$@SWCNT are obtained from DFT method. We verify that such encapsulation is noncovalent functionalization, and examined binding energies and charge transfers of $MCp_2$@(16,0)SWCNT systems. Consistent with recent experimental findings, the optimal distance between $FeCp_2$ center and near tube-wall is 4.7 (5.1) Å for the configuration where $MCp_2$'s five-fold axis is parallel (vertical) to nanotube axis, while the minimal diameter is 9.4 (10.2) Å to exothermically encapsulate $FeCp_2$ molecules. Finally we clarify the doping effects near the bandgap by encapsulations of $CoCp_2$ and $NiCp_2$.

**Introduction**

Carbon nanotubes (CNTs), which exhibit uniquely excellent structural, electronic and mechanical properties, have evoked a lot of research interest and been playing an important role in the fields of nanoscience and nanotechnology.[1-5] The surface of CNTs provides an attractive organic system that can interact with various molecules, atoms and nanostructures. Considerable works have demonstrated that covalent or noncovalent functionalization of CNTs surface can offer a new avenue to obtain p- and n-type semiconductor materials, which is important for constructing the complementary electronics.[6-12] In many respects, noncovalent functionalzation is more advantageous compared with the covalent one. For example, the noncovalent functionalization can immobilize target molecules on the sidewall of nanotubes without destroying the geometric structures of the tubes.[13, 14] Noncovalent functionalization of several organic molecules can also be a route for the separation of nanotubes based on chiral selectivity, owing to the charge transfer between the nanotubes and target molecules.[15-16]

On the other hand, due to their one-dimensional geometry, CNTs also provide a unique space for functionalization by encapsulation of dopants inside their hollow core. Until now, much research interest is focused on the encapsulated CNTs.[17-33] It has been experimentally shown that fullerenes and endohedral metallfullerenes can be inserted into CNTs, forming a peapod-like structure.[15, 18] Encapsulated organic/CNTs compounds are synthesized and characterized as well.[15, 17, 20, 21] Theoretical and experimental research results reveal that encapsulation of CNTs with organic molecules and fullerene can achieve p- and n-type dopings, based on the electron affinity and ionization energy of dopants.[20] Recently, endohedral functionalizations of CNTs with halogen atom[22, 23], metal atom[24-29] and metal carbide[30] were studied, from which insights into the modified electronic properties of CNTs have been attained. As the dopants occupy the inner space of CNTs, the doping state is rather stable in air. Remarkably, previous investigations have also predicted the stability of cobaltocene encapsulated inside CNTs.[19, 20, 31, 32] For instance, the experimental studies by Li and co-workers have shown that encapsulation of cobaltocene in carbon nanotubes are stable for nanotubes diameter of 9.4 Å.[19] They also found a charge transfer from the molecule to the nanotube wall due to the formation of localized impurity states, which could make these systems specially suitable to produce n-doped nanotubes. Encapsulation of ferrocene in CNTs was also reported by Guan and co-workers, who showed that ferrocene encapsulated CNTs could be simply synthesized and controllably doped.[31] On the theoretical side, the structures, stabilities and electronic properties of encapsulated CNTs with metallocene have been reported.[20, 33, 34] However, systematic investigations of encapsulated CNTs with various metallocenes in CNTs of different tube diameters are not yet available.

The objective of the present work is to investigate the functionalization of single-walled carbon nanotubes (SWCNTs) by encapsulation of dicyclopentadienyl metal ($MCp_2$, M=Fe, Co, Ni or metallocene), and to examine the possibility of obtaining p- or n-type semiconducting materials. In metallocene molecules, two π-electron systems $C_5H_5$ are bonded to the center metal, forming a sandwich structure. The most significant feature of the reactions with metallocene in general involves the aromatic nature, which is further expected to interact with CNTs. Using the density function theory (DFT) calculations, we have obtained the most energetically stable encapsulation geometries. For zigzag (16, 0) SWCNTs we mapped the electronic charge transfers from molecule to nanotube, and examined the binding energies of the encapsulated systems of (16, 0) SWCNTs with $MCp_2$ (labeled as $MCp_2$@(16, 0) SWCNT) with consideration of molecule's adiabatic ionization potential (AIP). Further total energy results reveal the optimal distance between the center of metallocene molecule and the near wall of tube, and the minimum SWCNT diameter required to exothermically encapsulate a metallocene molecule for different inserted structures. Finally, we also briefly discuss the electronic properties of the doped SWCNT by $MCp_2$ encapsulation.

**Method and model**

We carried out the *ab initio* DFT calculations using the spin-polarized local density approximation (LDA) with Perdew and Wang functional[35], as well as generalized-gradient approximation (GGA) with Perdew-Burke-Ernzerhof (PBE) functional[36] for comparison, and the double numerical basis set including polarization function (DNP basis set) implemented in the $DMol^3$ package[37]. The global orbital cutoff of basis sets is set to be 4.2Å in real space. The choice of the LDA is not fortuitous and it is more suitable than the GGA to study weakly interacting systems such as the $sp^2$-like materials. As indicated by the previous theoretical study, although the binding energies were slightly overestimated, the geometrical structures, charge distributions and electric spectra calculated by LDA were better than GGA.[21, 38] Zigzag single-wall carbon nanotubes were chosen as the benchmark model. This type of SWCNTs can exhibit diversified metallic and semiconducting characteristics. The periodic boundary condition was used with a tetragonal supercell of 40 Å × 40 Å × 8.439 Å, which consists of one $MCp_2$ in two periodic unit-cells of a zigzag SWCNT. The Brillouin zone was sampled using the Monkhorst-Pack scheme[38] by $1\times1\times2$ and $1\times1\times3$ special *k*-points for geometry optimization and energy calculations, respectively. The convergence thresholds were set as 0.00005 Ha in energy and 0.004 Ha/Å in force.

The binding energy ($E_b$) of functionalization SWCNT is defined as $E_b = E[\text{SWCNT@MCp}_2] - E[\text{SWCNT}] - E[\text{MCp}_2]$, where $E[\text{SWCNT@MCp}_2]$ is the energy of the MCp$_2$@SWCNT system, $E[\text{SWCNT}]$ is the total energy of pristine SWCNT, and $E[\text{MCp}_2]$ is the energy of an isolated MCp$_2$ molecule.[40,41] In such a definition, negative binding energy means exothermic encapsulation into SWCNT. The charge transfer between SWCNT and MCp$_2$ molecule is analyzed based on Mulliken method. In order to predict the electronic properties, we also performed DFT calculations on the band structures and density of states (DOS) using denser $1 \times 1 \times 12$ special $k$-points.

**Results and discussion**

First, we searched for the most stable configuration of the (16, 0) SWCNTs with endohedral MCp$_2$ doping. We chose (16, 0) SWCNT (diameter $d$=12.53 Å) as a representative zigzag SWCNT, which provides enough space to encapsulate MCp$_2$ in its hollow structure. Two intercalation configurations of MCp$_2$@(16, 0) SWCNT were considered and displayed in Figure 1, where the five-fold axis of MCp2 is parallel (labeled as Ip) or vertical to the tube axis (labeled as Iv). Full geometry optimization has been performed for the organometallic molecules and (16, 0) SWCNT separately prior to the static calculation of the MCp$_2$@(16, 0) SWCNT composite systems. Figure 2 displays the binding energy curves computed for the configurations depicted in Figure 1. The near distances ($D_n$) between the central metal atom of MCp$_2$ and the tube wall range from 4.27 to 6.27 Å, corresponding to the far distances ($D_f$) from 8.26 to 6.26 Å.

As shown in Figure 2(a), there is a noticeable difference between the two calculation methods (DFT-LDA and DFT-GGA) for the same compound configurations. For Ip configuration, the binging energy of FeCp$_2$@ (16, 0) SWCNT systems is between -0.30 and -0.65eV as a function of $D_n$, from the DFT-LDA calculation. The equilibrium distance of $D_n$ is 4.67 Å, where the iron atom deviates a distance of 1.60 Å away from the tube axis. Compared with FeCp$_2$, the equilibrium $D_n$ of CoCp$_2$ and NiCp$_2$ endohedral (16, 0) SWCNT is slightly shorter (4.57 Å), and the distance from center of the metallocenes to tube axis is about 1.70 Å. Remarkably, the equilibrium $D_n$ of 4.57 Å from DFT-LDA is in good agreement with the experimental value of nanotube radius for stable encapsulated CNTs with cobaltocene.[19] The equilibrium configuration of CoCp$_2$@(16, 0) SWCNT is more exothermic than that of FeCp$_2$@(16, 0) SWCNT and NiCp$_2$@(16, 0) SWCNT by 0.56 and 0.36 eV, suggesting a predicted stability in order of FeCp$_2$@SWCNT < NiCp$_2$@SWCNT < CoCp$_2$@SWCNT. In great contrast, the calculations by DFT-GGA underestimate the binding energies and overestimate the equilibrium distance $D_n$. For example, when FeCp$_2$ is encapsulated inside (16, 0) SWCNT, the GGA equilibrium $D_n$ is about 0.6 Å larger than that of LDA calculation and the binding energy at equilibrium site is about 0.5 eV bigger as well. Moreover, the GGA calculations show that all considered MCp$_2$@(16, 0) SWCNT systems have approximately indistinguishable equilibrium $D_n$ of 5.27 Å, and the metal atom keeps a distance of 1.00 Å away from the tube axis. Our calculated equilibrium distance of CoCp$_2$@(16, 0) SWCNT is agreement with the previous DFT-GGA investigation[20]. Similarly, the binding energy decreases by the same order of FeCp$_2$@SWCNT < NiCp$_2$@SWCNT < CoCp$_2$@SWCNT, based on the DFT-GGA calculations.

For Iv configurations, equilibrium distances $D_n$ between the central metal atom of MCp$_2$ and nanotube wall are larger than the Ip configurations. For FeCp$_2$, the equilibrium $D_n$ is 5.07 Å, exceeding the Ip configuration by 0.4 Å. CoCp$_2$ has the same equilibrium distance with FeCp$_2$, while NiCp$_2$ holds the relatively longer equilibrium distance about 5.27 Å. Likewise, CoCp$_2$@(16, 0) SWCNT gives the largest binding energy, NiCp$_2$@(16, 0) SWCNT system has a moderate $E_b$, and $E_b$ of doped (16, 0) SWCNT by FeCp$_2$ encapsulation is the lowest among the three metallocene compounds. However, Ip and Iv configurations provided binding energies $E_b$ close to each other, when the metallocene molecules are located at equilibrium sites. For CoCp$_2$, the difference of $E_b$ between the Ip configuration and Iv configuration is the largest, but less than 0.10 eV. These results indicate that the orientations of metallocene's fivefold axis inside the tube seem not to influence the binding energies, in turn, the stability of doped systems[19].

Based on the fully relaxed structures that the metallocene molecules are located at equilibrium sites, the optimized binding energies ($E_b$) and charge transfer (Q) between the nanotubes and metalloncenes are summarized in Table 1. Compared to initial configurations, the nanotube wall and the structures of metallocene molecules are almost unchanged upon encapsulation (shown in Supporting information). Calculations show that there is significant charge transfer from metallocene molecules to nanotube of about 0.164, 0.576 and 0355 e (0.111, 0.513 and 0.276 e) for the case of FeCp$_2$, CoCp$_2$ and NiCp$_2$ in Ip (Iv) configurations respectively. The direction of such charge transfers is again consistent with the experimental investigation.[19] It is obvious that the largest charger transfer correlates with the largest binding energy, which partially explains the binding energy order FeCp2@SWCNT < NiCp2@SWCNT < CoCp2@SWCNT. It is well known that the charger transfer of these doped SWCNTs change with the electron affinity (EA) or ionization potential (IP) of the encapsulated organic molecules[20]. So, we calculate the IP of these metalloncenes using Perdew-Wang functional and DNP basis sets, shown in Table 1. The calculated adiabatic IP's are 6.97, 5.13 and 5.94 eV for FeCp$_2$, CoCp$_2$ and NiCp$_2$, respectively. Therefore, CoCp$_2$ encapsulated inside SWCNT is expected to donate more electrons than $FeCp_2$ and $NiCp_2$, due to the relatively low AIP. As a consequence, $CoCp_2$@SWCNT has a larger binding energy than $FeCp_2$@SWCNT and $NiCp_2$@SWCNT.

In order to identify the nanotube diameter effect in detail, we examine a series of zigzag (n, 0) SWCNTs with endohedral metallocene. Additionally, these results could reveal the minimum SWCNT diameter required to exothermically encapsulate a metallocene molecule[19]. Only $FeCp_2$@SWCNT systems in Ip and Iv configurations are considered. The diameters of candidate SWCNTs from (10, 0) to (18, 0) range between 7.92 and 14.1 Å. Using the same approach, we first calculate the static total energy along different distance $D_n$ to search the equilibrium distance, and then fully optimize the structure with equilibrium distance. The binding energy curves computed for a $FeCp_2$ molecule inside all considered SWCNT are shown in Figure3, while the binding energies and charge transfer (Q) of the optimized structures are listed in Table 2.

As shown in Figure 3 (a), the equilibrium distances $D_n$ are about 4.70 Å for (18, 0) to (14, 0) SWCNT in Ip configurations, as well as for (12, 0) SWCNT in which Fe atom is just located at the tube axis. Therefore, it is reasonable to postulate that the optimal distance of $D_n$ is 4.70 Å, when $FeCp_2$ has been encapsulated into SWCNTs. However, for (13, 0) SWCNT in Ip configuration, the equilibrium distance is about 0.3 Å larger than 4.7 Å. Furthermore, for (11, 0) and (10, 0) SWCNTs, the distance between the center of metalloncenes and the tubes axis is greatly suppressed from 4.7 Å, due to little space in the interior of tubes (the radii of (11, 0) and (10, 0) are 4.30 and 3.96 Å, respectively). After full optimizations, we found that the nanotube walls are almost unchanged upon encapsulation of the $FeCp_2$ molecules for (18, 0) to (12, 0) SWCNTs (with large enough diameter), but they are distorted for (11, 0) and (10, 0) SWCNTs (diameter less than 9 Å).

The degree of ellipsoidal deformation from the circular tube cross section is $r_{max}/r_{min}$ = 7%. The results in Table 2 show that the optimized binding energies of encapsulated systems decrease with the decrease of tube's diameter from (18, 0) to (12, 0) SWCNTs with the lowest value of -1.28 eV, and then abruptly increase for (11, 0) and (10, 0) SWCNTs. This is explainable in terms of the distance ($D_f$) between the center of FeCp$_2$ and the far wall of tubes. As mentioned above, for SWCNTs from (18, 0) to (12, 0) the equilibrium $D_n$ is generally around 4.70 Å, thus as the tube diameters increase $D_f$ also increases. Therefore the interaction between encapsulated molecule and the far wall of tubes, which can additionally enjoy stabilization, gradually becomes weaker and results in increase of binding energies. However, for (13, 0) SWCNT the equilibrium $D_n$ is about 5.0 rather than 4.7 Å, again possibly because of the contribution from additional stabilization induced by $D_f$ of 5.2 Å is larger than the contribution of energy difference from change of $D_n$. For (12, 0) SWCNT (with diameter of 9.40 Å), the $D_n$ and $D_f$ are just equally 4.70 Å where the central Fe atom of FeCp$_2$ aligns with the SWCNT axis. This nanotube diameter of 9.40 Å for the most stable FeCp$_2$-encapsulated CNTs is also similar to the experimental finding for CoCp$_2$-encapsulated CNTs.[19] When the diameters are smaller than 9.40 Å for (11, 0) and (10, 0) SWCNTs, the interaction between encapsulated FeCp$_2$ and tube's wall change rapidly into repulsion, inducing dramatic binding energy increase even up to positive (endothermic) values. For Iv configurations, we find similar behaviors as the Ip case. For example, there is an optimal $D_n$ of about 5.1 Å (longer than that for Ip configurations by 0.4 Å); and the binding energies also decrease as the diameters of SWCNTs decrease, when the SWCNTs have larger diameters. However, the exception happens at (14, 0) SWCNT instead of (13, 0) SWCNT, where the equilibrium $D_n$ is 5.5 Å with a 0.4 Å deviation from the optimal distance; and the lowest binging energy can be obtained at (13, 0) SWCNT rather than (12, 0), when the diameter of (13, 0) SWCNT makes $D_n$ and $D_f$ of FeCp$_2$@(13, 0) SWCNT equal to the optimal distance (5.1 Å). In the same way, for SWCNTs with smaller diameter, the instability of encapsulated systems is abruptly enhanced. This behavior is also consistent with the above description for Ip configurations.

The computational results thus show that there exists an optimal distance from the center of metalocenes to the near tube's wall (4.70 and 5.10 Å for Ip and Iv configurations, respectively), and the minimum CNT diameter is 9.40 Å (10.20 Å) for Ip (Iv) configuration to exothermically encapsulate a metalocene molecule. This provides an excellent agreement with the diameter-selective encapsulation experimental results[19]. However, why is $D_n$ 4.7 Å for Ip configuration, but 5.10 Å for Iv? To answer this, we analyzed the distance from the corresponding carbon atoms of cyclopentadienyl to the near wall of nanotube. It is remarkable that the distances are about 3.5 Å for both configurations (Ip and Iv), i.e. a typical distance that an organic molecule is physically adsorbed onto the surface of CNTs. Therefore, we conclude that these capsulations are in fact physisorption, and they do not distort the nanotube walls.

Because of the significant charge transfer between the tube and intercalated molecules, we expect that the encapsulation of metallocenes inside carbon nanotubes could lead to some interesting effects on the electronic properties of SWCNTs[19]. The band structures and density of states (DOS) of pristine (16, 0) SWCNT, as well as stable Ip configurations after metallocene encapsulation, are calculated and plotted in Figure 4 and Figure 5. (16, 0) SWCNT is semiconducting with a DFT band gap of 0.50 eV as shown in Figures 4 (a) and 5 (a), in agreement with the theoretical studies reported in the latest literature, indicating that the model and method used in our paper are appropriate for the following analysis.

Compared with the pristine (16, 0) SWCNT, the encapsulations with $MCp_2$ all lead to downshift of the conduction band and valence band without much change of the original band gap of nanotube. And new states derived from the highest occupied molecular orbital (HOMO) of $MCp_2$ appear due to the charge transfer from $MCp_2$ to tubes[24]. As displayed in Figures 4 (b) and 5 (b), intercalation of $FeCp_2$ into SWCNT simply gives rise to a completely filled states at the valence band edge (VBE) and no influence to the electronic properties of SWCNT. Thus, encapsulation modification of SWCNT with $FeCp_2$ can be viewed as some kind of "harmless modification". Furthermore, $FeCp_2$ is paramagnetic because its core atom has a closed-shell electronic structure[25]. The local DOS of $FeCp_2$ molecule shown in Figure 5 (b) illustrates the new states derived from the HOMO state of the $FeCp_2$ molecule. The HOMO state can not drill through the VBE of nanotube and subsequently influence the electronic properties of nanotube around Fermi level. However, calculations based on the spin-polarized DFT (shown in Figures 4 (c)-(d) and 5 (c)-(d)) reveal that capsulation of $CoCp_2$ or $NiCp_2$ inside SWCNT can influence the electronic structures of SWCNT. For $CoCp_2$, in the spin-up band structure, two half-filled states appear near the conduction band edge (CBE), as well as within the spin-down conduction band away from the CBE. The results of local DOS of $CoCp_2$ (Figure 5 (c)) show that these half-filled states are induced by the singly occupied molecular orbital (SOMO) or HOMO of $CoCp_2$ molecule, indicating the SOMO states drill through the VBE of nanotube and reach to the tube's conduction band. Remarkably, we find that the Fermi level is located in conduction bands due to the downshift of CBE and the spin-up half-filled state from the SOMO of $CoCp_2$ is just below the Fermi level, confirming that the doped SWCNT exhibits n-type characteristics by encapsulation of $CoCp_2$ found in experiments[19]. On the other hand, when SWCNT is doped with $NiCp_2$ the capsulation results in the creation of spin-up half-filled state also reaching out above the VBE but within the band gap of SWCNT. The other state appears in conduction band even farther away above the CBE, as shown in the spin-down Figures 4 (d) and 5 (d). Therefore, the half-filled state is mainly contributed by the SOMO of $NiCp_2$ and is located just below the Fermi level in the band gap, predicting that the doped SWCNT displays a degenerate n-type behavior by encapsulation of $NiCp_2$.

On the basis of the results discussed above, it is reasonable to postulate that the capsulation of $MCp_2$ into SWCNTs could preserve or influence the electronic properties of SWCNTs. For $FeCp_2$, due to little charge transfer the molecular orbital of $FeCp_2$ can not drill through the VBE of tube, so the electronic property of SWCNTs has been largely unchanged. On the Contrary, for $CoCp_2$ or $NiCp_2$ with large charge transfer the unoccupied molecular orbitals drill through the VBE of tube and reach to the tube's conduction band or within SWCNTs' band gap. Therefore it induces the doped systems to exhibit n-type or degenerate n-type characteristics, with controlled electronic property change of CNTs. These localized impurity donor states below the conduction band have been experimentally observed to modify the emission and absorption processes[19]. As a result, the $MCp_2$@SWCNT systems possess the great potential for use in future nanoscale devices, without changing the exterior structure of nanotubes.

**Conclusion**

In summary, using density functional theory we have theoretically investigated the structural, energetic and electronic properties of doped SWCNTs by encapsulation of MCp$_2$ (M = Fe, Co, Ni). Two intercalation configurations (Ip and Iv) are considered, and we find that the binding energy changes as a function of the $D_n$. For the encapsulations of a given SWCNT, the stability and charge transfer increase as the AIP of metallocene decreases (CoCp$_2$ < NiCp$_2$ <FeCp$_2$). Through examining a series of zigzag (n, 0) SWCNTs encapsulated with metallocenes, we have found an optimal $D_n$ of 4.70 Å (5.10 Å) for Ip (Iv) configurations, and the minimum CNT diameter is 9.40 Å (10.20 Å) for Ip (Iv) configuration to exothermically encapsulate a metalocene molecule. The band structures and DOS computations verify that the encapsulation by metallocenes molecules (especially by CoCp$_2$ and NiCp$_2$) is an effective method to modulate the electronic properties of SWCNTs, along with the advantages in air stability and unchanged structure of nanotubes.

**ACKNOWLEDGMENT.** We gratefully acknowledge the National Natural Science Foundation of China (20673149, 20973204) and Guangdong Provincial Natural Science Foundation (9351027501000003). This work was partially sponsored by the highperformance grid computing platform of Sun Yat-sen University.

**Table 1.** Binding energies $E_b$, equilibrium $D_n$ and Mulliken charge transfer Q of MCp$_2$ (M = Fe, Co, Ni) molecules inside (16, 0) SWCNT. The calculated adiabatic ionization potentials (AIP's) are listed for metallocene molecules.

|    |                    | $D_n$ (Å) | $E_b$ (eV) | Q (e) | AIP (eV) |
|----|--------------------|-----------|------------|-------|----------|
| Ip | FeCp$_2$@(16, 0)   | 4.67      | -0.65      | 0.164 | 6.97     |
|    | CoCp$_2$@(16, 0)   | 4.57      | -1.21      | 0.576 | 5.13     |
|    | NiCp$_2$@(16, 0)   | 4.57      | -0.85      | 0.355 | 5.94     |
| Iv | FeCp$_2$@(16, 0)   | 5.07      | -0.65      | 0.111 | 6.97     |
|    | CoCp$_2$@(16, 0)   | 5.07      | -1.12      | 0.513 | 5.13     |
|    | NiCp$_2$@(16, 0)   | 5.27      | -0.79      | 0.276 | 5.94     |

**Table 2.** Binding energy $E_b$, equilibrium $D_n$ ($D_f$) and Mulliken charge transfer Q of FeCp$_2$ molecules inside SWCNTs from (18, 0) to (10, 0).

|         | Ip         |            |            |         | Iv         |            |            |         |
|---------|------------|------------|------------|---------|------------|------------|------------|---------|
|         | $D_n$ (Å)  | $D_f$ (Å)  | $E_b$ (eV) | Q (e)   | $D_n$ (Å)  | $D_f$ (Å)  | $E_b$ (eV) | Q (e)   |
| (18, 0) | 4.68       | 9.45       | -0.61      | 0.169   | 5.05       | 9.05       | -0.64      | 0.109   |
| (16, 0) | 4.67       | 7.87       | -0.65      | 0.164   | 5.07       | 7.47       | -0.65      | 0.111   |
| (15, 0) | 4.66       | 7.12       | -0.71      | 0.176   | 5.12       | 6.62       | -0.74      | 0.110   |
| (14, 0) | 4.73       | 6.23       | -0.83      | 0.172   | 5.48       | 5.48       | -1.04      | 0.117   |
| (13, 0) | 4.99       | 5.19       | -1.11      | 0.188   | 5.09       | 5.09       | -1.10      | 0.150   |
| (12, 0) | 4.70       | 4.70       | -1.28      | 0.234   | 4.70       | 4.70       | -0.00      | 0.190   |
| (11, 0) | 4.30       | 4.30       | -0.00      | 0.253   | 4.30       | 4.30       | 0.93       | 0.168   |
| (10, 0) | 3.96       | 3.96       | 1.18       | 0.214   | 3.96       | 3.96       | 1.14       | 0.214   |

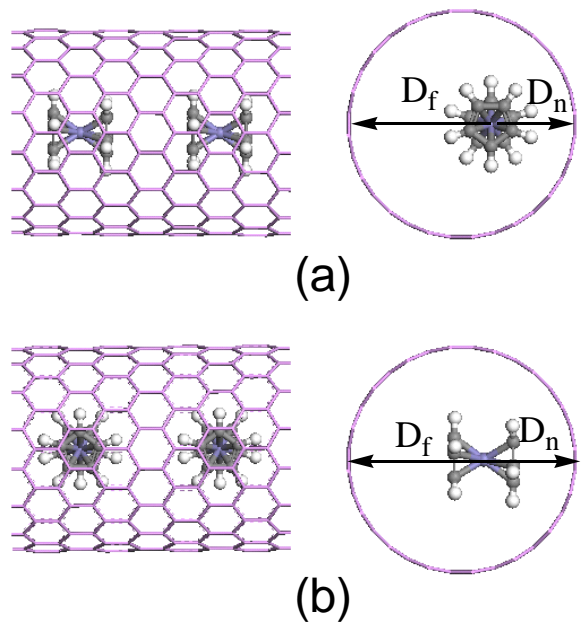

**Figure 1** Geometric structures of MCp$_2$ (M = Fe, Co, Ni) molecule encapsulated in (16, 0) SWCNT: (a) Ip configuration, (b) Iv configuration.

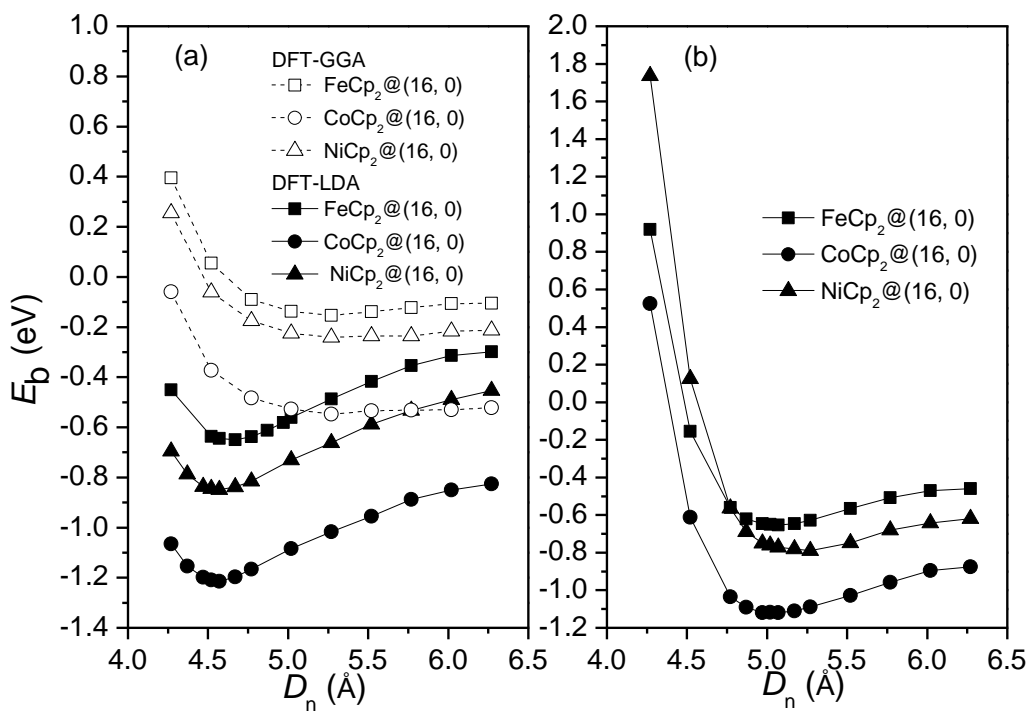

**Figure 2** Binding energies $E_b$ of MCp$_2$ (M = Fe, Co, Ni) molecule encapsulated (16, 0) SWCNT as a function of the $D_n$: (a) for Ip configuration (at LDA and GGA levels), (b) for Iv configuration (only at LDA level).

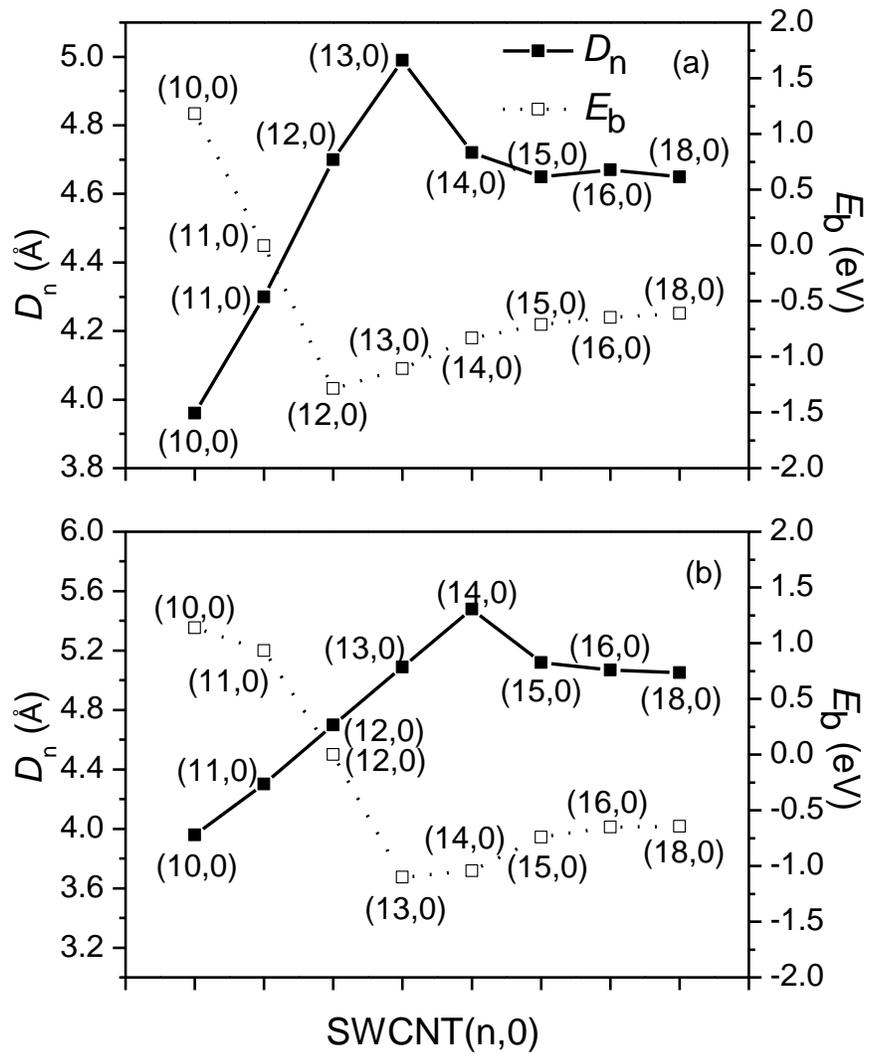

**Figure 3** Equilibrium $D_n$ and binding energies $E_b$ of FeCp$_2$ molecule encapsulated SWCNTs as a function of the SWCNTs: (a) for Ip configuration, (b) for Iv configuration.

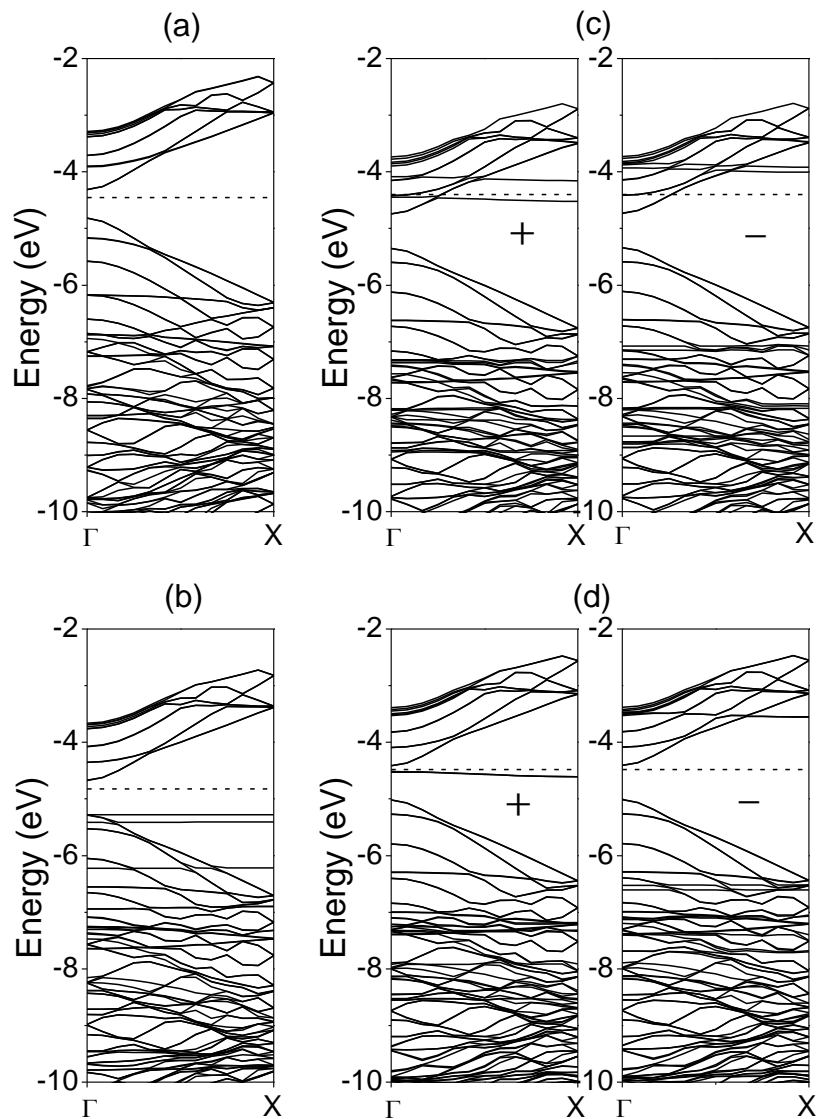

**Figure 4** Band structures of (a) pristine (16,0) SWCNT, (b) FeCp$_2$@(16,0)SWCNT, (c) CoCp$_2$@(16, 0)SWCNT, (d) NiCp$_2$@(16,0)SWCNT. The spin-up and spin-down electronic structures are distinguished with "+" and "-". Fermi level is indicated with a black dotted line.

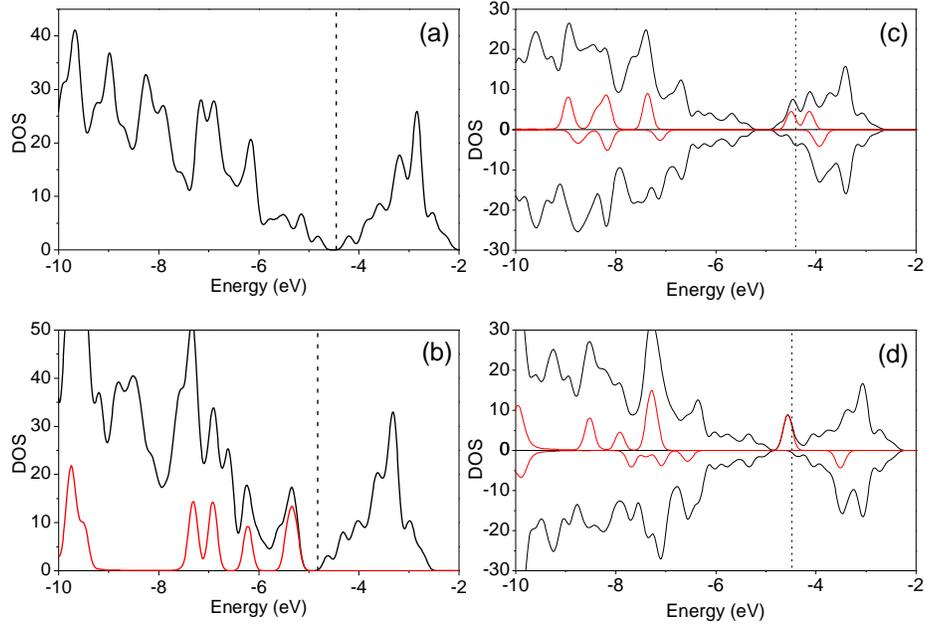

**Figure 5.** Density of states (DOS) of (a) pristine (16, 0) SWCNT, (b) FeCp$_2$@(16, 0)SWCNT, (c) CoCp$_2$@(16, 0)SWCNT, and (d) NiCp$_2$@(16, 0)SWCNT. Fermi level is indicated with a black dotted line. The projected DOS of MCp$_2$ is plotted with red lines.